\documentclass[
pra, 
aps,
twocolumn,
superscriptaddress,
longbibliography,
floatfix,
notitlepage]{revtex4-2}

\usepackage[utf8]{inputenc}
\usepackage[english]{babel}
\usepackage[T1]{fontenc}
\usepackage{lmodern}

\usepackage{amsmath,amsfonts,amssymb,amsthm,bm,times,dcolumn}

\usepackage{microtype}
\usepackage{braket}
\usepackage{gensymb} 
\usepackage{physics}

\usepackage[colorlinks={true}, citecolor={blue}, filecolor={blue}, 
linkcolor={blue}, urlcolor={blue}]{hyperref}
\usepackage{graphicx,color}
\usepackage[caption=false]{subfig}

\usepackage{soul}
\usepackage[normalem]{ulem}

\begin{document}
	
\preprint{APS/123-QED}

\title{Expanding a 4-qubit Dicke State to a 5-qubit Dicke State with Limited Qubit Access}

\author{Bibhuti Thapa}
\affiliation{Tokyo International University, 4-42-31 Higashi-Ikebukuro, Toshima-ku, Tokyo, 170-0013, Japan}

\author{Oberon Moran}
\affiliation{Tokyo International University, 4-42-31 Higashi-Ikebukuro, Toshima-ku, Tokyo, 170-0013, Japan}

\author{Duc-Kha Vu}
\affiliation{Tokyo International University, 4-42-31 Higashi-Ikebukuro, Toshima-ku, Tokyo, 170-0013, Japan}

\author{Fatih Ozaydin}
\email{mansursah@gmail.com}
\affiliation{Tokyo International University, 4-42-31 Higashi-Ikebukuro, Toshima-ku, Tokyo, 170-0013, Japan}
\affiliation{Nanoelectronics Research Center, Kosuyolu Mah., Lambaci Sok., Kosuyolu Sit., No:9E/3  Kadikoy, Istanbul, 34718, T\"urkiye}
	
\date{\today}

\begin{abstract}
In scenarios where full access to all qubits of a multipartite quantum system is available and global operations can be implemented, the preparation of arbitrary entangled states is theoretically straightforward. However, practical constraints often limit direct control over all qubits. In this work, we first present an efficient method for preparing a four-qubit Dicke state, and then demonstrate how a four-qubit Dicke state can be expanded to a five-qubit Dicke state even when only a subset of qubits is accessible. We propose a quantum circuit that achieves this transformation under restricted control, and support our analytical derivation with numerical simulations. We further carry out a robustness analysis of our circuit under imperfect gate implementations and find that it retains high fidelity for experimentally relevant levels of coherent over-rotation errors, confirming its resilience to realistic noise.
\end{abstract}

\maketitle

\section{Introduction}\label{sec:introduction}

Multipartite entangled states are essential resources for realizing a wide range of quantum information processing tasks, including measurement-based quantum computation~\cite{raussendorf2001one,briegel2009measurement}, quantum metrology~\cite{ma2011quantum,ozaydin2014phase,song2012quantum}, entanglement-assisted quantum error correction~\cite{wang2019entanglement, wilde2007entanglement}, entanglement swapping across quantum networks~\cite{su2016quantum,ottaviani2019multipartite}, quantum heat engines~\cite{dag2019temperature,ozaydin2024engineering}, distributed quantum computing~\cite{cacciapuoti2019quantum,cuomo2020towards}, threshold quantum cryptography~\cite{tokunaga2005threshold}, and quantum blockchain and consensus protocols~\cite{li2022efficient}.

Despite their central role, the preparation of multipartite entangled states remains a significant challenge, particularly when the entanglement structure is complex or when constraints limit access to qubits or control operations. Even in the ideal case where all qubits are accessible and global operations are allowed on arbitrary subsets, generating certain classes of entangled states can be nontrivial. While GHZ states~\cite{nielsen2010quantum}, as well as unweighted~\cite{raussendorf2001one} and even weighted graph states~\cite{tame2009compact}, can be constructed relatively straightforwardly, for other classes such as $W$ states and their generalizations—namely Dicke states—their generation continues to be an active area of research due to the more intricate correlations involved.

Counterdiabatic driving has been explored as a route to Dicke state preparation. In particular, Ref.~\cite{opatrny2016counterdiabatic} demonstrated that adiabatically transforming a spin coherent state into a maximally spin-squeezed Dicke state can be accelerated using compensating operators, thereby suppressing unwanted diabatic transitions.

Recent advances also explore the use of simultaneous multi-qubit entangling operations to reduce circuit depth and improve fidelity in near-term devices. For example, Gu et al.~\cite{gu2021fast} proposed constructing multiqubit gates by applying multiple two-qubit gates in parallel, enabling fast preparation of large entangled states such as Dicke and GHZ states without requiring hardware modifications.

Alternative approaches exploit global control techniques for nondeterministic Dicke state generation. For instance, the method proposed in~\cite{wang2021preparing} employs a phase-estimation-based protocol with collective $ZZ$ interactions and ancilla qubit measurements, offering a scalable route for applications such as magnetic sensing with spin ensembles.

Another line of research focuses on algorithmic approaches for constructing arbitrary Dicke states. For instance, Chakraborty et al.~\cite{chakraborty2014efficient} proposed efficient quantum algorithms that leverage symmetric Boolean functions, Krawtchouk polynomials, and elements from the Deutsch–Jozsa and Grover algorithms, as well as parity measurements and biased Hadamard transformations, to enable flexible Dicke state generation.

While not directly proposing a Dicke state preparation protocol, related progress has been made in large-scale trapped-ion systems. For example, Ref.~\cite{obvsil2019multipath} demonstrated optical multipath interference from strings of up to 53 trapped ions, achieving high spatial indistinguishability and coherence preservation. In their conclusions, the authors anticipate that these results could be directly extended to the optical generation of large atomic Dicke states, which, with addressed qubit operations, could be transformed into a broad spectrum of Dicke states for diverse applications.

Circuits for the deterministic preparation of small-size $W$~\cite{yesilyurt2015optical} and Dicke states~\cite{bartschi2019deterministic} have been proposed without any constraints on the availability of global operations.

Since a major challenge in experimental quantum computing is to minimize the number of costly two-qubit controlled gates, the design and optimization of quantum circuits for preparing Dicke states has been an active area of investigation. For example, it has been shown that a four-qubit Dicke state can be prepared by implementing 16 controlled-NOT (CNOT) gates~\cite{dutta2024following}, and this number was subsequently reduced to 12 CNOT gates through circuit simplification~\cite{mukherjee2020preparing}. More recently, by adopting a divide-and-conquer strategy that decomposes the preparation into smaller entanglement-building subroutines, the CNOT count was further lowered to 7~\cite{aktar2022divide}. These advancements highlight the ongoing effort to improve resource efficiency in Dicke state generation, as reducing the two-qubit gate overhead directly translates to higher experimental feasibility by mitigating decoherence, gate errors, and overall circuit execution time.

However, in scenarios where operations are restricted to a limited subset of qubits, several strategies have been developed to overcome these challenges. In particular, fusion~\cite{ozdemir2011optical,bugu2013enhancing,yesilyurt2013optical,ozaydin2014fusing,bugu2020,li2016generating,CavityRefs2}, expansion~\cite{ozaydin2021deterministic,yesilyurt2016deterministic,zang2016deterministic}, and collision-based~\cite{ccakmak2019robust} approaches have been utilized to construct larger $W$ states under restricted qubit access.

While mathematical descriptions of quantum operators capable of transforming Dicke states have been studied~\cite{kobayashi2014universal}, to the best of our knowledge, there has been no explicit circuit design that implements such transformations under limited qubit access. 

In this work, we address this gap by first proposing a resource-efficient quantum circuit for the deterministic preparation of a four-qubit Dicke state using only six two-qubit controlled gates, which, to the best of our knowledge, represents the lowest reported count for such a construction. Building on this compact design, we then introduce a protocol that expands a four-qubit Dicke state into a five-qubit Dicke state while operating under a restricted-access constraint, where only a subset of qubits is available for manipulation. We analyze both the success and failure probabilities of the transformation and, in the latter case, investigate the possibility of recycling the remnant entangled state to improve overall resource efficiency. Our approach is validated through extensive numerical simulations over 100{,}000 runs, and we further perform a robustness analysis under coherent over-rotation errors in the multi-qubit gates, demonstrating that the protocol maintains high fidelity for realistic experimental error levels.

\section{Preliminaries and the Model}\label{sec:Model} 
%

A $|W_n\rangle$ state is a special case of the more general $n$-qubit Dicke state~\cite{dicke1954coherence}, characterized by a single excitation, i.e., $|D_n^{(1)}\rangle$. The $n$-qubit, three-qubit, and four-qubit $W$ states can be expressed as particular instances of Dicke states~\cite{dur2000three,ozaydin2021deterministic}

\begin{eqnarray}
	|W_n\rangle &=& {1 \over \sqrt{n} } ( |0\rangle^{\otimes (n-1)} |1\rangle + \sqrt{n-1} | W_{n-1} \rangle |0 \rangle),\\
	|W_3\rangle &=&  \frac{ |001\rangle + |010\rangle +|100\rangle } {\sqrt{3}},\\	
	|W_4\rangle &=&  \frac{ |0001\rangle + |0010\rangle +|0100\rangle +|1000\rangle} {\sqrt{4}}.
\end{eqnarray}

A $|\overline{W}_n\rangle$ state is an $n$-qubit Dicke state with $n-1$ excitations, i.e. $|D_n^{(n-1)}\rangle$.
While two-qubit GHZ and $W$ states both reduce to Bell states, i.e., $\frac{|00\rangle + |11\rangle}{\sqrt{2}}$ for GHZ-type and $\frac{|01\rangle + |10\rangle}{\sqrt{2}}$ for $W$-type, the structure of Dicke states becomes more nuanced as the number of qubits increases. For three qubits, the Dicke state with a single excitation corresponds to the standard $|W_3\rangle$ state, whereas the Dicke state with two excitations corresponds to a state denoted $|\overline{W}_3\rangle$. The latter can be obtained by applying a local Pauli-X operation to each qubit of the $|W_3\rangle$ state. Consequently, three-qubit Dicke states are equivalent to $W$ states up to local operations and classical communication (LOCC)~\cite{nielsen2010quantum}. 

However, $n$-qubit Dicke states with $n>3$ and excitation numbers other than $1$ or $n-1$ are not equivalent to $n$-qubit $W$ states under LOCC, and their increased structural complexity necessitates special consideration for their preparation. 
In the following, we demonstrate how $|D_4^{(2)}\rangle$, the four-qubit Dicke state with two excitations

\begin{eqnarray}\label{eq:Dicke42}
	|D_4^{(2)}\rangle & = &\frac{1}{\sqrt{6}} ( |1100\rangle + |1010\rangle + |1001\rangle \\  \nonumber
	& & + |0110\rangle + |0101\rangle + |0011\rangle )
\end{eqnarray}

\noindent 
can be expanded into $|D_5^{(3)}\rangle$, the five-qubit Dicke state with three excitations.

\begin{eqnarray}\label{eq:Dicke53}
	|D_5^{(3)}\rangle\! & = &\!\! \frac{1}{\sqrt{10}} (
	|11100\rangle + |11010\rangle + |11001\rangle + |10110\rangle \\ \nonumber
	& & + |10101\rangle + |10011\rangle + |01110\rangle + |01101\rangle \\ \nonumber  
	& & + |01011\rangle + |00111\rangle ).
\end{eqnarray}

\section{Expansion Strategies}\label{sec:ExpansionStrategies}

Before expanding the four-qubit Dicke state \( |D_4^{(2)}\rangle \) into the five-qubit Dicke state \( |D_5^{(3)}\rangle \), we first demonstrate how \( |D_4^{(2)}\rangle \) can be constructed deterministically from the three-qubit $W$ state, \( |W_3\rangle \). This intermediate step both motivates our overall protocol and illustrates the structured buildup of Dicke states from smaller entangled resources.
We will also use this step for the \textit{recycling} process, later, in the case of a failure in the actual \( |D_4^{(2)}\rangle \rightarrow |D_5^{(3)}\rangle \) transformation.

\subsection{Expanding a three-qubit $W$ state into a four-qubit Dicke state}\label{sec:WtoDicke}

We adopt the following notational conventions: the Hadamard gate is denoted by $H$, the Pauli-X (NOT) gate by $X$, and a single-qubit gate $G$ applied to qubit $i$ is written as $G^{i}$. For multiqubit controlled operations, we denote the gate $G$ with control qubits $\{c_1, c_2, \dots, c_n\}$ and target qubit $t$ as $G^{c_1, c_2, \dots, c_n; t}$.

\begin{figure}[t!]
	\includegraphics[width=0.75\linewidth]{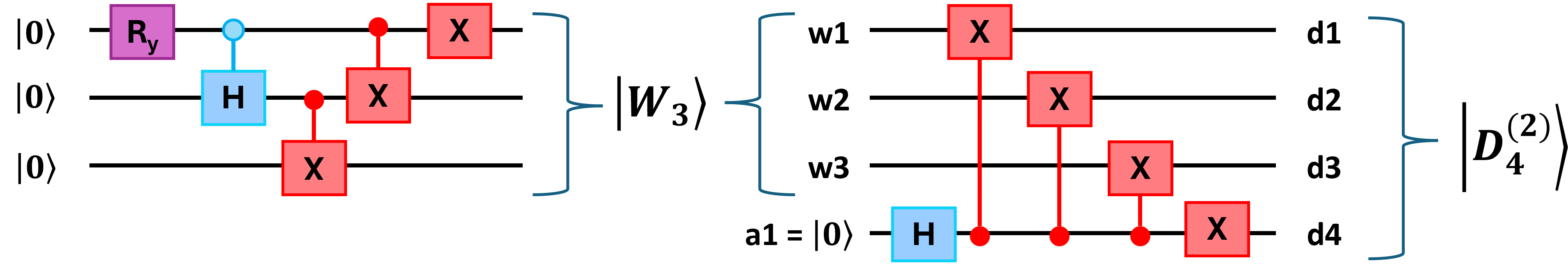}
	\caption{Left. Circuit for deterministic preparation of a $|W^3\rangle$ state from three separable qubits. Right. Circuit for deterministic expansion of a three-qubit $W$ state into a four-qubit Dicke state.\label{fig:fig1}}
\end{figure}

As illustrated in Fig.~\ref{fig:fig1}, we begin with a $|W_3\rangle$ state with qubits $\{w1, w2, w3\}$, which itself can be efficiently prepared using known fusion~\cite{bugu2013enhancing} or expansion~\cite{yesilyurt2016deterministic} techniques. To this state, we append an ancillary qubit $a1$ initialized in the computational basis state $|0\rangle$. We then apply a carefully designed sequence of single- and multi-qubit gates that transform the combined four-qubit system into the Dicke state \( |D_4^{(2)}\rangle \) with qubits $\{d1, d2, d3, d4\}$. The entire transformation is deterministic and requires access to all four qubits during the operation. This preparatory stage forms the basis for the subsequent expansion to five qubits under restricted-access conditions, and demonstrates that Dicke states can be hierarchically constructed from simpler entangled states as

\onecolumngrid
\onecolumngrid
\twocolumngrid
\twocolumngrid

\begin{eqnarray}\label{eq:WtoDicke}
	& & |W_3\rangle \otimes |0\rangle \\ \nonumber
	& \rightarrow & X^4 \ \text{CNOT}^{4;3} \ \text{CNOT}^{4;2} \ \text{CNOT}^{4;1} \ H^4 (|W_3\rangle \otimes |0\rangle) \\
	& = & |D_4^{(2)}\rangle. \nonumber
\end{eqnarray}

Requiring only six two-qubit controlled gates, our proposed circuit achieves, to the best of our knowledge, the lowest two-qubit gate count reported for the deterministic preparation of a four-qubit Dicke state. This represents an improvement over existing designs, including those in~\cite{dutta2024following,mukherjee2020preparing,aktar2022divide}, where the most efficient approach still requires a minimum of seven two-qubit controlled gates. Given that two-qubit operations are typically the most error-prone and time-consuming elements in quantum circuits, this reduction directly enhances the experimental feasibility of our method by lowering cumulative gate errors, reducing decoherence effects, and shortening the overall circuit depth. Consequently, our design not only advances the state of the art in Dicke state preparation but also provides a practical advantage for near-term quantum hardware implementations.

\subsection{Expanding a four-qubit Dicke state into a five-qubit Dicke state}\label{sec:DicketoDicke}

Unlike the approach presented in~\cite{bartschi2019deterministic}, where full access to all qubits is assumed throughout the transformation process, our protocol introduces an important constraint: only three out of the four qubits in the initial Dicke state are accessible for quantum operations, while one qubit remains untouched. This constraint models a more realistic setting often encountered in distributed or modular quantum computing architectures, where direct access to every qubit in a multipartite entangled system may not be available due to physical separation, hardware limitations, or network topology restrictions.

Following Ref.~\cite{kobayashi2014universal}, we first define $k$, the number of accessible qubits as

\begin{equation}	
	k \geq 
	\begin{cases}
		M_1 & \text{for } m_0 > 0 \\
		M_0 & \text{for } m_1 > 0
	\end{cases}
\end{equation}

\noindent
where $M_0$  and $M_1$ are the number of qubits in states $|0\rangle$ and $|1\rangle$, respectively, and $m_0$ and $m_1$  are the number of qubits in states $|0\rangle$ and $|1\rangle$ added for expansion of the circuit, respectively.

We choose the number of accessible qubits $k=3$. To apply the transformation, we derive in the Appendix the decomposition of the initial Dicke state in terms of its accessible and inaccessible subsystems, A and B, respectively, as

\begin{equation}
	\label{eq:D42_decomposition}
	\begin{aligned}
		|D_4^2\rangle_{AB} &= 
		\sqrt{\frac{1}{2}} |D_3^2\rangle_A |0\rangle_B
		+ \sqrt{\frac{1}{2}} |D_3^1\rangle_A |1\rangle_B,
	\end{aligned}
\end{equation}

\noindent
and the target state as

\begin{equation}
	\label{eq:D53_decomposition}
	\begin{aligned}
		|D_5^3\rangle_{AB} &=  \sqrt{\frac{2}{5}} |D_4^3\rangle_A |0\rangle_B
		+ \sqrt{\frac{3}{5}} |D_4^2\rangle_A |1\rangle_B.
	\end{aligned}
\end{equation}\\

We calculate $p_{\max}$, the maximum probability of success to be obtained with this model as 

\begin{equation}
	\begin{aligned}
		q_j &= \frac{\binom{k}{M_1-j}}{\binom{k+n}{M_1+m_1-j}} 
		&&\text{for } j = 0, 1, \\[0.5ex]
		q_j &= \frac{\binom{3}{2-j}}{\binom{4}{3-j}}, \\[0.5ex]
		q_{\min} &= \min\left\{\frac{3}{4}, \frac{1}{2}\right\} = \frac{1}{2}, \\[0.5ex]
		p_{\max} &= q_{\min} \frac{\binom{5}{3}}{\binom{4}{2}} 
		= \frac{1}{2} \cdot \frac{10}{6} 
		= \frac{5}{6}.
	\end{aligned}
	\label{eq:pmax}
\end{equation}

\onecolumngrid
\onecolumngrid
\begin{figure}[h!]
	\includegraphics[width=1\linewidth]{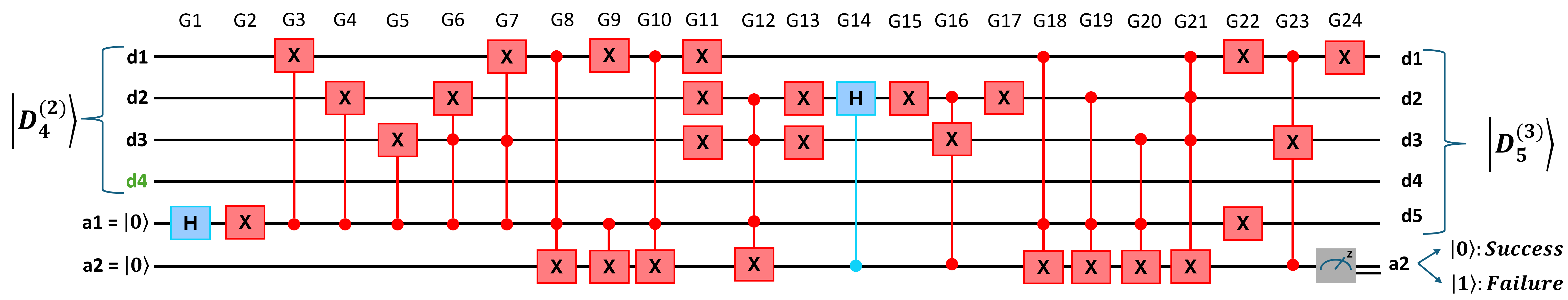}
	\caption{Quantum circuit implementing the transformation \( |D_4^{(2)}\rangle \otimes |0\rangle \otimes |0\rangle \rightarrow |D_5^{(3)}\rangle \) under restricted qubit access. The input consists of four qubits \( \{d1, d2, d3, d4\} \) prepared in the Dicke state \( |D_4^{(2)}\rangle \), and two ancillary qubits \( a1 \) and \( a2 \), both initialized in the state \( |0\rangle \). The circuit employs single-qubit gates \( H \) (Hadamard) and \( X \) (Pauli-X), along with multi-qubit gates including CNOT, CH (controlled-Hadamard), CCNOT (Toffoli), and CCCNOT gates. All operations are applied exclusively to qubits \( \{d1, d2, d3, a1, a2\} \), while qubit \( d4 \), shown in green, remains completely untouched throughout the entire process. After the gate sequence, the flag qubit \( a2 \) is measured in the computational basis. If the outcome is \( |0\rangle \), the remaining five qubits form the desired Dicke state \( |D_5^{(3)}\rangle \), with the ancillary qubit \( a1 \) successfully integrated into the entangled system.\label{fig:fig3}}
\end{figure}
\twocolumngrid
\twocolumngrid

\begin{figure}[h!]
	\includegraphics[width=0.9\linewidth]{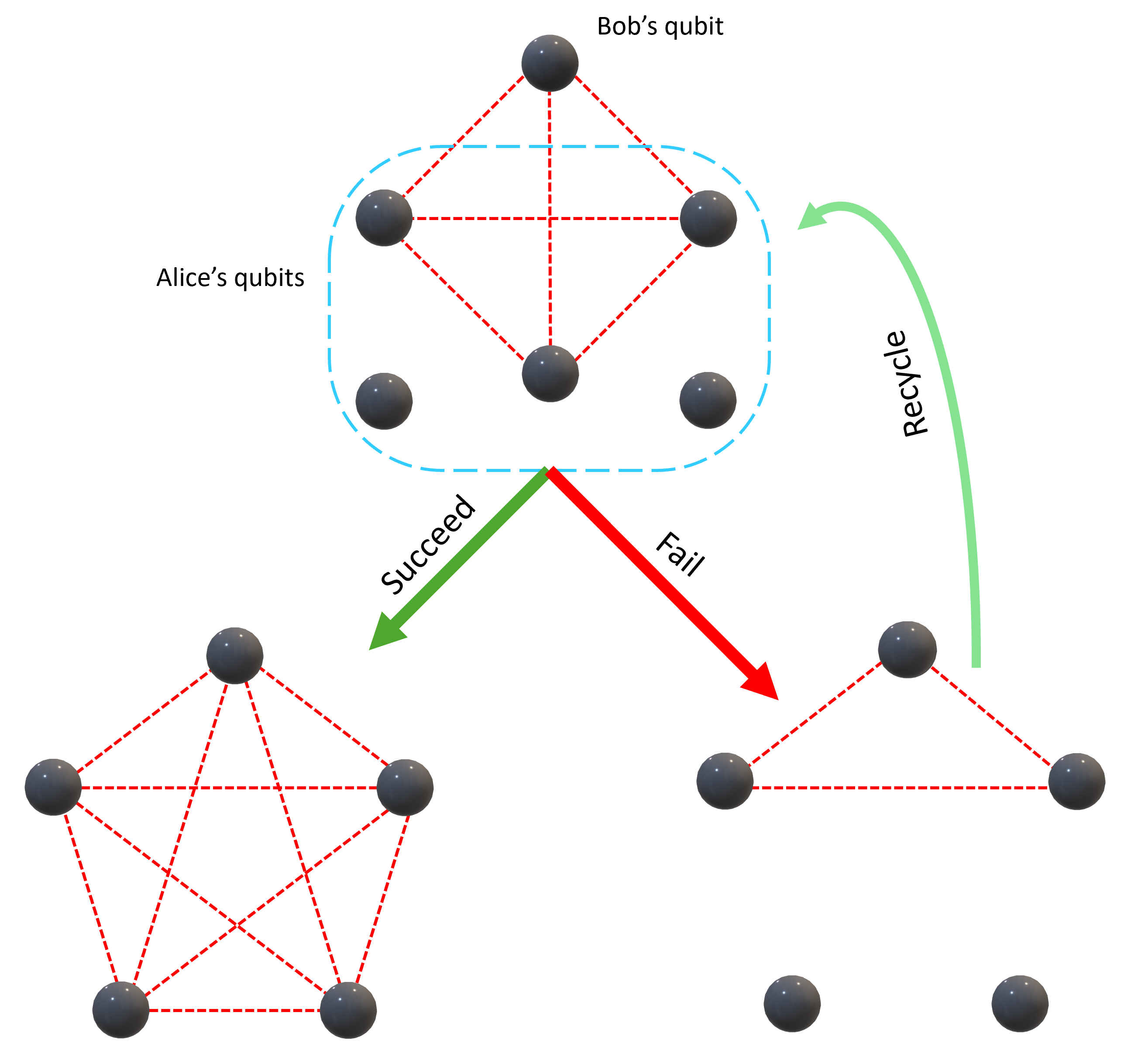}
	\caption{Illustration of the Dicke state expansion strategy under restricted qubit access. Rather than assuming global access to all qubits, we consider a scenario in which one of the qubits—denoted as being on Bob's site—is completely inaccessible during the operation. Alice, who has access to the remaining three qubits of the initial \( |D_4^{(2)}\rangle \) state, introduces two ancillary qubits initialized in the \( |0\rangle \) state and performs all gate operations locally on her side. After applying a sequence of single- and multi-qubit gates, Alice measures the second ancillary qubit in the computational basis. If the measurement outcome is \( |0\rangle \), the expansion is considered successful, and the five-qubit system—comprising Alice’s three data qubits, the previously inaccessible qubit on Bob's side, and the first ancilla—forms the target Dicke state \( |D_5^{(3)}\rangle \).  If the measurement outcome is instead \( |1\rangle \), the expansion attempt is deemed unsuccessful. However, this is not a catastrophic failure: the process results in the loss of one qubit from the original Dicke state, yielding a three-qubit state that is locally equivalent to a \( W \)-like state. This remaining entangled state can be recycled to reconstruct a new \( |D_4^{(2)}\rangle \) using the method described in Section~\ref{sec:WtoDicke}, and thus the outcome is better described as a \textit{recyclable} state rather than a complete failure.
		\label{fig:fig2}}
\end{figure}

To design the circuit for implementing the above model, our strategy begins with a four-qubit Dicke state \( |D_4^{(2)}\rangle \) defined over qubits \( \{d1, d2, d3, d4\} \), and introduces two ancillary qubits \( \{a1, a2\} \) initialized in the \( |0\rangle \) state. The key constraint in our protocol is that all operations—both single- and multi-qubit gates—are performed without interacting with qubit \( d4 \), which remains entirely inaccessible during the process. As illustrated in Fig.~\ref{fig:fig2}, this situation can be conceptually described by assigning qubits \( \{d1, d2, d3\} \) and the ancillae \( \{a1, a2\} \) to Alice, and the inaccessible qubit \( d4 \) to Bob. Alice performs all gate operations on her local qubits, after which she measures the ancillary qubit \( a2 \), which can now be considered a \textit{flag} qubit, in the computational basis.

The measurement outcome of \( a2 \) determines the success or failure of the expansion process. With a success probability of \( p_s = \frac{5}{6} \), the measurement yields the \( |0\rangle \) state, indicating that the five qubits \( \{d1, d2, d3, d4, a1\} \) are now collectively in the Dicke state \( |D_5^{(3)}\rangle \). In this successful case, the ancillary qubit \( a1 \) effectively joins the entangled Dicke state, and we relabel it as qubit \( d5 \) to reflect its incorporation into the system. Conversely, with failure probability \( p_f = \frac{1}{6} \), the measurement yields the \( |1\rangle \) outcome, signifying that the expansion attempt was unsuccessful. However, this outcome does not fully destroy the entanglement; instead, the original four-qubit Dicke state is reduced to the three-qubit Dicke-like state

\begin{equation}\label{eq:WLike}
	|\overline{W^L_3}\rangle = \frac{ |110 \rangle}{2} +\frac{ |101 \rangle}{2} + \frac{ |011 \rangle}{\sqrt{2}}.
\end{equation}

As discussed earlier, this state is equivalent to a three-qubit $W$-like state, which can serve as a resource for reconstructing a new \( |D_4^{(2)}\rangle \) state via the procedure described in Section~\ref{sec:WtoDicke}. In line with the terminology used in prior fusion-based protocols~\cite{ozdemir2011optical,bugu2013enhancing,ozaydin2014fusing}, we refer to this outcome as yielding a \textit{recyclable} state rather than a strict failure.
The fidelity of this $|\overline{W^L_3}\rangle$ state to a $|\overline{W_3}\rangle$ is equal to 0.97. Hence, the fact that it is not a genuine $W$ would have a minimal impact in the recycling process. Note that such $W$-like states are also required for the realization of many quantum tasks, including perfect teleportation and superdense coding~\cite{li2016generating}.

The quantum circuit used to implement the transformation \( |D_4^{(2)}\rangle \otimes |0\rangle \otimes |0\rangle \rightarrow |D_5^{(3)}\rangle \) is shown in Fig.~\ref{fig:fig3}. The circuit employs a combination of single-qubit gates—specifically the Hadamard (H) and Pauli-X (X) gates—as well as multi-qubit gates including CNOT (controlled-NOT), CH (controlled-Hadamard), CCNOT (controlled-controlled-NOT or Toffoli), and CCCNOT (controlled-controlled-controlled-NOT) operations. These gates are applied in a structured sequence designed to manipulate the system while preserving the symmetry and excitation number required for Dicke state construction.

In the circuit diagram, qubit \( d4 \), assumed to reside on Bob's side, is highlighted in green to emphasize that it remains completely untouched throughout the process. This visual distinction underlines a key constraint of our protocol: the entire transformation is achieved without performing any operation, measurement, or entangling gate on qubit \( d4 \), demonstrating the feasibility of Dicke state expansion under restricted access conditions.

It is worth noting that the success probability obtained from the circuit, success probability of $p_s = \frac{5}{6}$ achieves the maximum success probability $p_{max}$ derived mathematically in Eq.~\ref{eq:pmax}.  

In the Appendix, we present the proposed circuit in more detail with the description of each quantum gate.

\subsection{Simulation Results}\label{sec:SimulationResults}
\begin{figure}[t!]
	\includegraphics[width=1\linewidth]{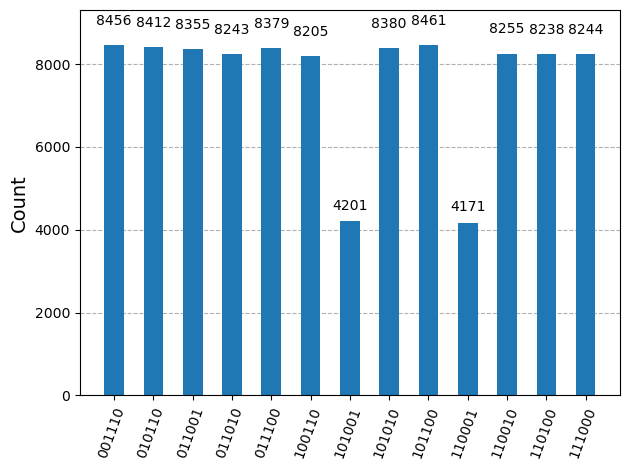}
	\caption{ 
		Simulated results of the quantum circuit in Fig.~\ref{fig:fig3} for expanding a \( |D_4^{(2)}\rangle \) state into a \( |D_5^{(3)}\rangle \) state over 100{,}000 runs. The $y$-axis shows the number of occurrences of each computational-basis outcome for all qubits after measurement. On the $x$-axis, the last qubit corresponds to the ancilla qubit \( a2 \), serving as the \textit{flag} qubit. If \( a2 \) is found in the \( |1\rangle \) state, the expansion attempt fails: the fourth and fifth qubits become separable, and the first three qubits collapse into the recyclable state \( |\overline{W^L_3}\rangle \) given in Eq.~\ref{eq:WLike}. Conversely, if \( a2 \) is measured in the \( |0\rangle \) state, the attempt succeeds, and the first five qubits form the Dicke state \( |D_5^{(3)}\rangle \). In the successful case, the frequency of each computational-basis term is approximately $\frac{5}{6}$, consistent with the expected statistics of an ideal Dicke state.
		\label{fig:fig4}}
\end{figure}

To evaluate the performance of our proposed protocol, we numerically simulated the quantum circuit shown in Fig.~\ref{fig:fig3} using 100{,}000 independent runs. The results are presented in Fig.~\ref{fig:fig4}, where the $y$-axis represents the number of occurrences for each possible computational-basis outcome of the qubits, and the $x$-axis enumerates the measured bit strings from left to right. The final qubit in each bit string corresponds to the ancilla qubit \( a2 \), which acts as a \textit{flag} qubit indicating the success or failure of the expansion attempt.

If the measurement outcome of \( a2 \) is \( |1\rangle \), the expansion attempt is considered unsuccessful. In this case, the fourth and fifth qubits become separable from the remaining register, and the first three qubits collapse into the recyclable state \( |\overline{W^L_3}\rangle \), as defined in Eq.~\ref{eq:WLike}. Although this outcome does not yield the target state, the resulting entangled three-qubit state can be reused to generate a fresh \( |D_4^{(2)}\rangle \) state, thereby improving overall resource efficiency.

On the other hand, if \( a2 \) is measured in the \( |0\rangle \) state, the expansion is successful: the first five qubits of the register form the target Dicke state \( |D_5^{(3)}\rangle \). The simulated output distribution in this case shows that each computational-basis term appears with a relative frequency close to $\frac{5}{6}$, in perfect agreement with the theoretical expectation for a perfectly prepared Dicke state. This almost-uniform distribution of basis states confirms that the circuit design preserves the symmetric structure of the Dicke state and validates the correctness of our protocol under ideal noise-free simulation.

\subsection{Robustness Analysis}
In any realistic quantum computing platform, multi-qubit gates such as controlled-NOT (CNOT), controlled-Hadamard (CH), and their multi-controlled generalizations are subject to imperfections due to calibration errors, control signal distortions, and decoherence during gate execution. One common form of coherent error in physical implementations is an \emph{over-rotation} or \emph{under-rotation} of the target qubit's rotation angle relative to its ideal value. In order to evaluate the robustness of our Dicke state expansion protocol against such coherent gate errors, we introduce a simple yet experimentally motivated error model parameterized by a deviation angle $\theta$.

\subsubsection{Error model}
In the ideal case ($\theta=0$), the target qubit operation $U_{\text{target}}$ in a controlled gate is implemented exactly as intended. In our error model, however, this operation is preceded by an extra coherent rotation around the $X$-axis of the Bloch sphere by an angle $\theta$, described by
\begin{equation}
	R_x(\theta) =
	\begin{pmatrix}
		\cos\left(\frac{\theta}{2}\right) & -i\sin\left(\frac{\theta}{2}\right) \\
		-i\sin\left(\frac{\theta}{2}\right) & \cos\left(\frac{\theta}{2}\right)
	\end{pmatrix}.
	\label{eq:Rx}
\end{equation}
The noisy target operation is therefore given by $R_x(\theta) U_{\text{target}}$. When $\theta=0$, Eq.~(\ref{eq:Rx}) reduces to the $2\times 2$ identity matrix and the ideal gate is recovered.

\subsubsection{Noisy CNOT and CH gates}
The ideal CNOT gate acting on two qubits (control = first, target = second) is given by
\begin{equation}
	\mathrm{CNOT} =
	\begin{pmatrix}
		1 & 0 & 0 & 0 \\
		0 & 1 & 0 & 0 \\
		0 & 0 & 0 & 1 \\
		0 & 0 & 1 & 0
	\end{pmatrix}.
	\label{eq:CNOTideal}
\end{equation}
In our model, the imperfect CNOT becomes
\begin{equation}
	\mathrm{CNOT}(\theta) =
	\left( |0\rangle\langle 0| \otimes I_2 \right)
	+ \left( |1\rangle\langle 1| \otimes R_x(\theta) X \right),
	\label{eq:CNOTtheta}
\end{equation}
where $X$ is the Pauli-$X$ matrix and $I_2$ is the $2\times 2$ identity matrix. For $\theta=0$, Eq.~(\ref{eq:CNOTtheta}) reduces to the ideal CNOT in Eq.~(\ref{eq:CNOTideal}).

Similarly, the ideal CH gate is
\begin{equation}
	\mathrm{CH} =
	\left( |0\rangle\langle 0| \otimes I_2 \right)
	+ \left( |1\rangle\langle 1| \otimes H \right),
	\label{eq:CHideal}
\end{equation}
where
\begin{equation}
	H = \frac{1}{\sqrt{2}}
	\begin{pmatrix}
		1 & 1 \\
		1 & -1
	\end{pmatrix}
	\label{eq:H}
\end{equation}
is the Hadamard matrix. In our model, the noisy CH becomes
\begin{equation}
	\mathrm{CH}(\theta) =
	\left( |0\rangle\langle 0| \otimes I_2 \right)
	+ \left( |1\rangle\langle 1| \otimes R_x(\theta) H \right),
	\label{eq:CHtheta}
\end{equation}
which again recovers the ideal gate when $\theta=0$.

It is important to note that the above over-rotation model is applied uniformly to \emph{all} controlled gates appearing in the circuit, regardless of the number or arrangement of control qubits or the specific identity of the target qubit. In each case, the imperfect gate is constructed by appending the extra $R_x(\theta)$ rotation on the target qubit immediately before the intended target operation, while the control subspace is enforced exactly as in the ideal case. This formulation ensures that the error model is consistent across CNOT, CH, CCNOT, CCCNOT, and any other multi-controlled gate, thereby providing a coherent and uniform framework for robustness analysis.

\subsubsection{Fidelity-based robustness metric}
To quantify the robustness of our Dicke state expansion protocol under coherent over-rotation errors, we compute the fidelity
\begin{equation}
	F(\theta) = \left| \langle \psi_{\mathrm{ideal}} | \psi_{\mathrm{noisy}}(\theta) \rangle \right|^2,
	\label{eq:Fidelity}
\end{equation}
where $|\psi_{\mathrm{ideal}}\rangle$ is the output state of the protocol under ideal gates, and $|\psi_{\mathrm{noisy}}(\theta)\rangle$ is the output under the imperfect gates of Eqs.~(\ref{eq:CNOTtheta}) and (\ref{eq:CHtheta}) for a given $\theta$. In the absence of error ($\theta=0$), we have $F(0) = 1$ by construction. As $\theta$ increases, coherent deviations accumulate across the sequence of multi-qubit gates, leading to a reduction in fidelity.

\subsubsection{Results}
\begin{figure}[t!]
	\centering
	\includegraphics[width=0.8\linewidth]{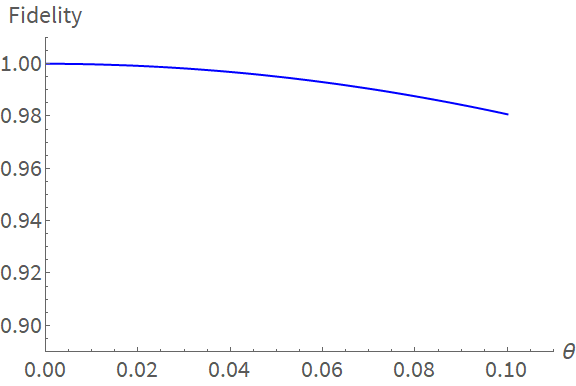}
	\caption{Fidelity $F(\theta)$ between the ideal Dicke state produced by our expansion protocol and the state produced under coherent over-rotation errors of magnitude $\theta$ in all multi-qubit gates. The fidelity remains above $0.99$ for $\theta \lesssim 0.01$ radians, indicating robustness against typical experimental imperfections.}
	\label{fig:DickePlot}
\end{figure}

The fidelity $F(\theta)$ for our Dicke state expansion circuit was numerically evaluated for $0 \leq \theta \leq 0.1$ radians, which is greater by an order than the typical over-rotation magnitudes reported in current superconducting qubit experiments~\cite{chen2016measuring}. The results are shown in Fig.~\ref{fig:DickePlot}, where we observe a slow decay of $F(\theta)$, consistent with perturbative error accumulation. These results confirm that our protocol maintains high fidelity for realistic values of $\theta$, highlighting its robustness to coherent gate imperfections.

\section{Discussions}
While the first proposed circuit  for preparing a $|D_4^{(2)}\rangle$ state achieves a reduction in the number of two-qubit controlled gates compared to previously reported approaches, we emphasize that we do not claim our second circuit to be the absolute optimal circuit for implementing the transformation $|D_4^{(2)}\rangle \rightarrow |D_5^{(3)}\rangle$. There may still be room for further optimization, particularly in reducing the total number of controlled operations without sacrificing fidelity. Our primary motivation in this work has been to design and demonstrate a scheme that performs the expansion while operating under a realistic constraint: only a subset of the qubits in the initial state is accessible for manipulation, with at least one qubit remaining completely untouched. This reflects practical scenarios in distributed or modular quantum architectures, where physical or architectural limitations prevent global access to all qubits.

The proposed approach is not limited to the specific case of expanding a four-qubit Dicke state to a five-qubit Dicke state. The design principles can, in principle, be adapted to richer scenarios where subgroups of qubits are inaccessible due to hardware constraints, communication latency, or isolation in separate modules. As a direction for future work, we plan to investigate more general strategies for preparing larger Dicke states under such access constraints, potentially combining expansion techniques with fusion-based protocols and optimization methods for gate reduction.

\section{Conclusions}
We have first presented a quantum circuit for expanding a three-qubit $W$ state into a four-qubit Dicke state with two excitations, $|D_4^{(2)}\rangle$. To our knowledge, the proposed method requires the fewest two-qubit controlled gates reported for a Dicke state expansion in this configuration, using only six such gates. Then we have presented a quantum circuit for expanding the  $|D_4^{(2)}\rangle$ state into a five-qubit Dicke state with three excitations, $|D_5^{(3)}\rangle$, under the constraint that one of the qubits in the initial state remains completely untouched. We analyzed both success and failure outcomes, demonstrating that the failure case yields a recyclable three-qubit $W$-lik state, which can be reused for subsequent Dicke state preparation attempts.
Our numerical simulations over $10^5$ runs confirm the correctness of the design, and our robustness analysis indicates that the circuit maintains high fidelity even in the presence of controlled-gate overrotation errors of realistic magnitude. These results suggest that the scheme is not only resource-efficient but also experimentally viable in near-term quantum processors, especially in distributed or modular settings. 

\section*{Acknowledgements}

F.O. acknowledges Tokyo International University Personal Research Fund.



\section*{Appendix}

	\subsection{Decomposition of Dicke States}
\subsubsection{The Initial State {$|D_4^2\rangle$}}
The summation bounds are:
\begin{align*}
	\alpha &= \max\{M_1 - k, 0\} = \max\{2 - 3, 0\} = 0, \\
	\beta &= \min\{N - k, M_1\} = \min\{4 - 3, 2\} = 1.
\end{align*}

\begin{equation}
	\label{eq:A_D42_decomposition}
	\begin{aligned}
		|D_4^2\rangle_{AB} &= \sum_{j=0}^{1} \sqrt{\frac{
				\binom{k}{M_1 - j} \binom{N-k}{j}
			}{\binom{N}{M_1}}} 
		|D_{k}^{M_1 - j}\rangle_A \, |D_{N-k}^{j}\rangle_B \\[1ex]
		&= \sum_{j=0}^{1} \sqrt{\frac{
				\binom{3}{2 - j} \binom{1}{j}
			}{\binom{4}{2}}} 
		|D_{3}^{2 - j}\rangle_A \, |D_{1}^{j}\rangle_B \\
		&= \sqrt{\frac{\binom{3}{2} \binom{1}{0}}{\binom{4}{2}}} |D_3^2\rangle_A |0\rangle_B
		+ \sqrt{\frac{\binom{3}{1} \binom{1}{1}}{\binom{4}{2}}} |D_3^1\rangle_A |1\rangle_B \\[1ex]
		&= \sqrt{\frac{3 \cdot 1}{6}} |D_3^2\rangle_A |0\rangle_B 
		+ \sqrt{\frac{3 \cdot 1}{6}} |D_3^1\rangle_A |1\rangle_B \\[1ex]
		&= \sqrt{\frac{1}{2}} |D_3^2\rangle_A |0\rangle_B 
		+ \sqrt{\frac{1}{2}} |D_3^1\rangle_A |1\rangle_B.
	\end{aligned}
\end{equation}

\subsubsection{The Target State $|D_5^3\rangle$}
The summation bounds are:
\begin{align*}
	\alpha' &= \max\{M_1 + m_1 - k -n, 0\} = \max\{-1, 0\} = 0, \\
	\beta' &= \min\{N - k, M_1 + m_1\} = \min\{1, 3\} = 1.
\end{align*}

\begin{equation}
	\label{eq:A_D53_decomposition}
	\begin{aligned}
		|D_5^3\rangle_{AB} &= \sum_{j=\alpha'}^{\beta'} \sqrt{\frac{
				\binom{k+n}{M_1+m_1-j} \binom{N-k}{j}
			}{\binom{N+n}{M_1+m_1}}} |D_{k+n}^{M_1 +m_1 - j}\rangle_A \, |D_{N-k}^{j}\rangle_B\\[1ex]
		&= \sum_{j=0}^{1} \sqrt{\frac{
				\binom{4}{3 - j} \binom{1}{j}
			}{\binom{5}{3}}} 
		|D_{4}^{3 - j}\rangle_A \, |D_{1}^{j}\rangle_B \\
		&= \sqrt{\frac{4 \cdot 1}{10}} |D_4^{3}\rangle_A |0\rangle_B + \sqrt{\frac{6 \cdot 1}{10}} |D_4^{2}\rangle_A |1\rangle_B \\[1ex]
		&= \sqrt{\frac{2}{5}} |D_4^3\rangle_A |0\rangle_B + \sqrt{\frac{3}{5}} |D_4^2\rangle_A |1\rangle_B.
	\end{aligned}
\end{equation}

	\subsection{Gate sequence}\label{sec:GateSequence}
	In this section, we detail our quantum circuit in Fig.~\ref{fig:fig3} labeling each gate, and present the gates through Eqs.A1-A23.
	\begin{eqnarray}
		G1 & = & H^{a1}, \\
		G2 & = & X^{a1}, \\
		G3 & = & \text{CNOT}^{a1; d1}, \\
		G4 & = & \text{CNOT}^{a1; d2}, \\
		G5 & = & \text{CNOT}^{a1; d3}, \\
		G6 & = & \text{CCNOT}^{a1,d3; d2}, \\
		G7 & = & \text{CCNOT}^{a1,d3; d1}, \\
		G8 & = & \text{CCNOT}^{d1,a1; a2}, \\
		G9 & = & X^{d1} \text{CNOT}^{a1; a2}, \\
		G10 & = & G8, \\
		G11 & = & X^{d1} X^{d2} X^{d3}, \\
		G12 & = & \text{CCCNOT}^{d1,d3,a1; a2}, \\
		G13 & = & X^{d2} X^{d3}, \\
		G14 & = & \text{CH}^{a2;d2}, \\
		G15 & = & X^{d2}, \\	
		G16 & = & \text{CCNOT}^{d2,a2; d3}, \\
		G17 & = & X^{d2}, \\
		G18 & = & G8 \\
		G19 & = & \text{CCNOT}^{d2,a1; a2}, \\
		G20 & = & \text{CCNOT}^{d3,a1; a2}, \\
		G21 & = & \text{CCCNOT}^{d1,d2,d3; a2}, \\
		G22 & = & X^{d1} X^{a1} \\
		G23 & = & \text{CCNOT}^{d1,a2; d3}, \\
		G24 & = & X^{d1}.
	\end{eqnarray}

\clearpage
\bibliography{C:/Users/seval/Dropbox/OQuL/OQuLBib/OQuL}

\end{document}